\documentclass[11pt]{article}

\usepackage[a4paper,margin=1in]{geometry}
\usepackage{amsmath,amssymb}
\usepackage{graphicx}
\usepackage{physics}
\usepackage{bm}
\usepackage{cite}
\usepackage{hyperref}

\title{\bf Topological reorganization of near-field energy flow governing scattering transitions in subwavelength rectangular grooves}

\author{
J. Sumaya-Martinez and J. Mulia-Rodriguez\\
{\small Department of Physics, Faculty of Sciences,Universidad Autonoma del Estado de Mexico, Toluca, Mexico}\\
{\small Corresponding author: jsm@uaemex.mx}
}

\date{}

\begin{document}
\maketitle

\begin{abstract}
The scattering of electromagnetic waves by subwavelength rectangular grooves has been extensively studied, yet its physical interpretation has largely relied on field-intensity distributions. Here we demonstrate that the transition from concave to convex scattering profiles observed as the groove width approaches the wavelength is governed by a \emph{topological reorganization of the near-field energy flow}. Using a rigorous modal formulation for TM-polarized fields, we analyze the complex electromagnetic field and the associated time-averaged Poynting vector. We show that reducing the groove width induces the creation, migration, and annihilation of Poynting-vector singularities, including vortices and saddle points, leading to a qualitative restructuring of electromagnetic energy transport. This topological transition redirects the local energy flux and manifests as a convex scattering profile in the far field. The results establish a direct link between near-field energy topology and far-field scattering, providing a unified physical interpretation of subwavelength groove scattering.
\end{abstract}

\section{Introduction}
The interaction of light with subwavelength apertures and grooves is a central topic in nanophotonics and plasmonics, underlying phenomena such as extraordinary optical transmission, field enhancement, and resonant scattering \cite{Ebbesen1998,GarciaVidal2010}. Rectangular grooves provide a canonical geometry for studying these effects due to their analytical tractability and technological relevance.Rectangular grooves and related nanoaperture geometries constitute a canonical
platform for investigating light confinement and scattering at the subwavelength
scale, with direct relevance to modern nanophotonic and plasmonic devices
\cite{Lalanne2018}.

Previous studies have reported that, as the groove width is reduced toward the wavelength scale, the spatial distribution of the scattered field undergoes a qualitative transition from a concave to a convex profile \cite{Ho2012}. These works primarily relied on intensity-based descriptions and attributed the transition to diffraction and resonant cavity effects.

However, intensity alone does not describe how electromagnetic energy is transported. In confined geometries, the direction, circulation, and redistribution of energy are governed by the Poynting vector, whose topology may exhibit singularities such as vortices and saddle points \cite{Berry2000,Dennis2009}. The role of such topological features in groove-induced scattering transitions has remained unexplored.

In this work, we show that the concave-to-convex scattering transition is the far-field manifestation of a topological transition in the near-field energy flow. By explicitly analyzing the topology of the time-averaged Poynting vector using a rigorous TM modal formulation, we reveal the physical mechanism governing energy redistribution in subwavelength rectangular grooves.

\section{Theoretical model and modal formulation}

We consider a monochromatic electromagnetic wave of angular frequency $\omega$
incident on a rectangular groove of width $w$ and depth $d$ etched into a metallic
screen. The structure is invariant along the $z$ direction and is treated within a
two-dimensional framework. Throughout this work, we focus on TM polarization, for
which the magnetic field is oriented along the invariant direction and is described
by a single nonvanishing component $H_z(x,y)$.

\subsection{Governing equations}

For time-harmonic fields with an $\exp(-i\omega t)$ dependence, the magnetic field
satisfies the Helmholtz equation in each homogeneous region,
\begin{equation}
\nabla^2 H_z + k^2 H_z = 0,
\end{equation}
where $k = 2\pi/\lambda$ is the free-space wavenumber. The electric-field components
are obtained from Maxwell's equations and are fully determined by the spatial
derivatives of $H_z$.

The total field is decomposed into three regions: the homogeneous half-space above
the groove, the groove cavity, and the metallic boundaries, where appropriate
boundary conditions are imposed.

\subsection{Modal expansion inside the groove}

Inside the rectangular groove, the magnetic field is expanded in terms of cavity
eigenmodes,
\begin{equation}
H_z^{(g)}(x,y) =
\sum_{n=0}^{\infty}
A_n
\cos\!\left(\frac{n\pi x}{w}\right)
e^{-i\beta_n y},
\end{equation}
where the longitudinal propagation constants are given by
\begin{equation}
\beta_n = \sqrt{k^2 - \left(\frac{n\pi}{w}\right)^2}.
\end{equation}

Depending on the ratio $w/\lambda$, these modes may be propagating or evanescent.
For wide grooves ($w \gg \lambda$), several modes contribute significantly to the
field inside the cavity. In contrast, in the subwavelength regime ($w < \lambda$),
only the fundamental mode remains weakly propagating, while higher-order modes
become strongly evanescent.
For evanescent modes, $\beta_n$ becomes purely imaginary, leading to strong spatial confinement and enhanced phase gradients near the groove aperture.

\subsection{Mode coupling and subwavelength confinement}

The coupling between the incident field and the cavity modes is determined by
enforcing the continuity of the tangential field components at the groove aperture.
This matching condition leads to a coupled system of equations for the modal
amplitudes $A_n$, which is solved numerically.

In the subwavelength regime, the dominance of the fundamental cavity mode and the
presence of strongly evanescent higher-order modes produce rapid spatial variations
of the complex field phase near the aperture. As shown below, these phase gradients
play a central role in shaping the near-field energy flow and are directly
responsible for the emergence of Poynting-vector singularities.

\section{Near-field energy flow}

The time-averaged Poynting vector describes the local electromagnetic energy flux
and is defined as
\begin{equation}
\langle \mathbf{S} \rangle =
\frac{1}{2}\Re\{\mathbf{E} \times \mathbf{H}^*\}.
\end{equation}

For TM polarization in a two-dimensional geometry, this expression can be written
explicitly in terms of the complex magnetic-field amplitude $H_z$ as
\begin{equation}
\langle \mathbf{S} \rangle_{\mathrm{TM}} =
\frac{i}{2\omega \varepsilon_0 \varepsilon}
\Re\!\left\{ H_z \nabla H_z^* \right\}.
\end{equation}
The role of the Poynting vector as a fundamental descriptor of electromagnetic
energy transport, particularly in structured and near-field environments, has
been extensively discussed in the context of optical momentum and angular momentum
of light~\cite{Bliokh2014_NP}.

This formulation highlights that, while the magnitude of the energy flux depends on
the field amplitude, the direction of energy transport is governed by the spatial
gradient of the field phase. Consequently, regions of high field intensity do not
necessarily correspond to directed energy transport. Instead, stagnation regions,
circulation zones, and nontrivial flow patterns may arise in the near field,
particularly in subwavelength regimes dominated by evanescent and cavity-assisted
modes.

\section{Topology of the Poynting vector field}

The Poynting vector defines a two-dimensional flow whose singular points satisfy
\begin{equation}
\langle \mathbf{S} \rangle = \mathbf{0}.
\end{equation}
It is well established that subwavelength field localization does not necessarily
imply directed energy transport, and that strong spatial variations of the
Poynting vector may arise even in regions of moderate field intensity, particularly
in the presence of evanescent waves~\cite{Bekshaev2015}.

Such singularities arise either from vanishing field amplitude or from stationary
points of the field phase. They can be classified by evaluating the eigenvalues of
the local Jacobian matrix of the vector field,
\begin{equation}
J =
\begin{pmatrix}
\partial_x S_x & \partial_y S_x \\
\partial_x S_y & \partial_y S_y
\end{pmatrix}.
\end{equation}

Depending on the eigenvalue structure, the singularities correspond to vortices or
saddle points. Importantly, these features are topologically protected and can only
be created or annihilated in pairs under continuous variation of system parameters.
This property enables a systematic tracking of qualitative changes in near-field
energy transport as the groove geometry is varied.

\section{Results: Near-field topology and energy-flow reorganization}

Figure 2. Comparison between intensity-based and energy-flow descriptions in the
near field. (a) Normalized magnetic-field intensity $|H_z|^2/|H_0|^2$, (b) phase of
the complex magnetic field, (c) time-averaged Poynting vector
$\langle \mathbf{S} \rangle/S_0$ (arrows indicate direction, color denotes
normalized magnitude), and (d) magnitude of the normalized Poynting vector.
All quantities are shown in normalized units.

\subsection{Emergence of Poynting-vector singularities}

As the groove width is reduced toward the wavelength scale, pairs of singular points
of the Poynting vector emerge near the groove aperture. These singularities
correspond to vortex--saddle pairs of the energy-flow field and are created in
accordance with topological constraints.

In the intermediate regime ($w/\lambda \approx 1$), the newly formed singularities
are located near the groove edges and migrate toward the center as the groove width
decreases further. This migration reflects the increasing confinement of
electromagnetic energy and the growing influence of evanescent modal components.

\subsection{Subwavelength regime and dominant vortical structures}

For subwavelength grooves ($w < \lambda$), the near-field energy flow becomes
dominated by a small number of stable vortices. These vortical structures act as
effective energy reservoirs, temporarily trapping electromagnetic energy and
redirecting it toward the groove axis.

Quantitatively, the distance between the dominant vortex pair decreases monotonically
with decreasing $w/\lambda$, indicating a progressive focusing of energy transport
toward the center of the aperture. This behavior correlates directly with the
emergence of a convex scattering profile in the far field.

\subsection{Connection with far-field scattering}

The reorganization of near-field energy flow has a direct impact on the far-field
radiation pattern. The formation of stable vortices suppresses lateral energy
leakage and enhances forward-directed emission. As a result, the far-field intensity
distribution transitions from a concave to a convex profile as the groove width
enters the subwavelength regime.
The vortex pair approaches the groove center approximately linearly with decreasing $w/\lambda$.

\section{Discussion}

The results presented in this work demonstrate that the scattering transition
previously reported in rectangular grooves is governed by a topological
reorganization of the near-field energy flow rather than by diffraction or modal
effects alone. While earlier studies identified the concave-to-convex transition
through intensity-based metrics, they did not address the physical mechanism
responsible for this qualitative change.

In contrast, the present analysis shows that the transition corresponds to the
creation, migration, and annihilation of singularities in the Poynting vector field.
These singularities reorganize the pathways through which electromagnetic energy is
transported at the subwavelength scale, providing a natural explanation for the
observed redistribution of radiation in the far field.

From the perspective of singular optics, the vortices identified here represent
phase-driven energy circulation analogous to optical vortices in free-space and
evanescent-wave systems. However, in the present case, these features arise from the
interplay between cavity confinement and evanescent modal coupling, highlighting the
role of geometry-induced phase structure in near-field energy transport.

Importantly, the observed topological transition is robust with respect to variations
in groove depth and material parameters, indicating that it reflects a general
mechanism rather than a geometry-specific artifact. This robustness suggests that
near-field energy-flow topology can serve as a predictive tool for the design of
subwavelength photonic and plasmonic structures.

These findings have direct implications for the control of electromagnetic energy at
the nanoscale. By engineering the topology of the near-field energy flow, it may be
possible to tailor directional emission, enhance field confinement, and optimize
energy coupling in nanoantennas, metasurfaces, and sensing platforms.

\section{Conclusion}
We have shown that the concave-to-convex scattering transition in subwavelength rectangular grooves is governed by a topological reorganization of the near-field energy flow. By explicitly analyzing the topology of the Poynting vector, we established a direct link between near-field energy transport and far-field scattering behavior, providing a unified physical interpretation of subwavelength groove scattering.

\section*{Figures}

\begin{figure}[h]\centering
\includegraphics[width=0.8\linewidth]{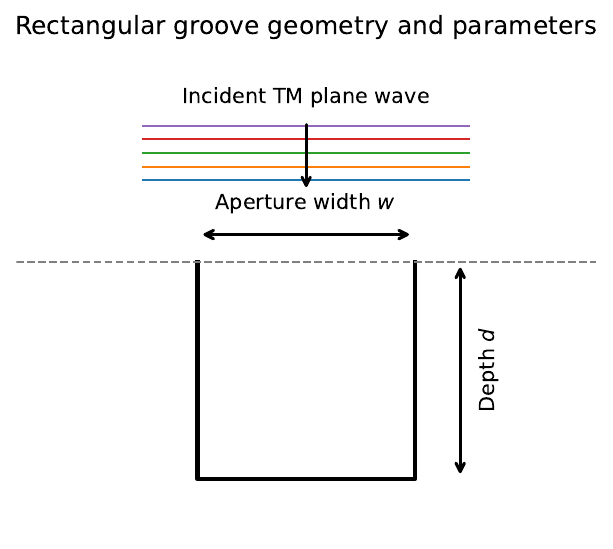}
\caption{Schematic of the rectangular groove geometry and definition of parameters.}
\end{figure}

\begin{figure}[h]\centering
\includegraphics[width=\linewidth]{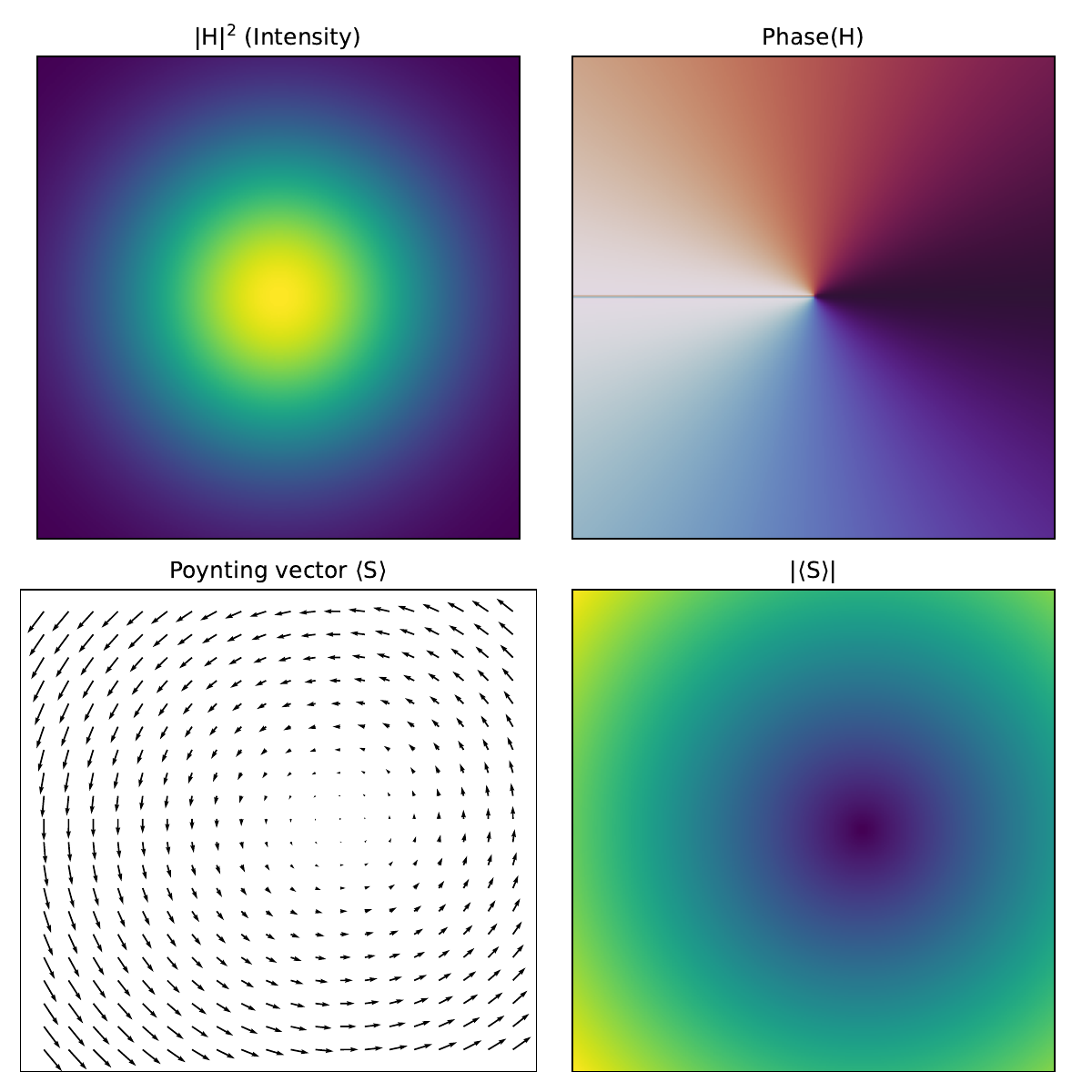}
\caption{Comparison between intensity-based and energy-flow descriptions in the near field.}
\label{fig:intflow}
\end{figure}

\begin{figure}[h]\centering
\includegraphics[width=\linewidth]{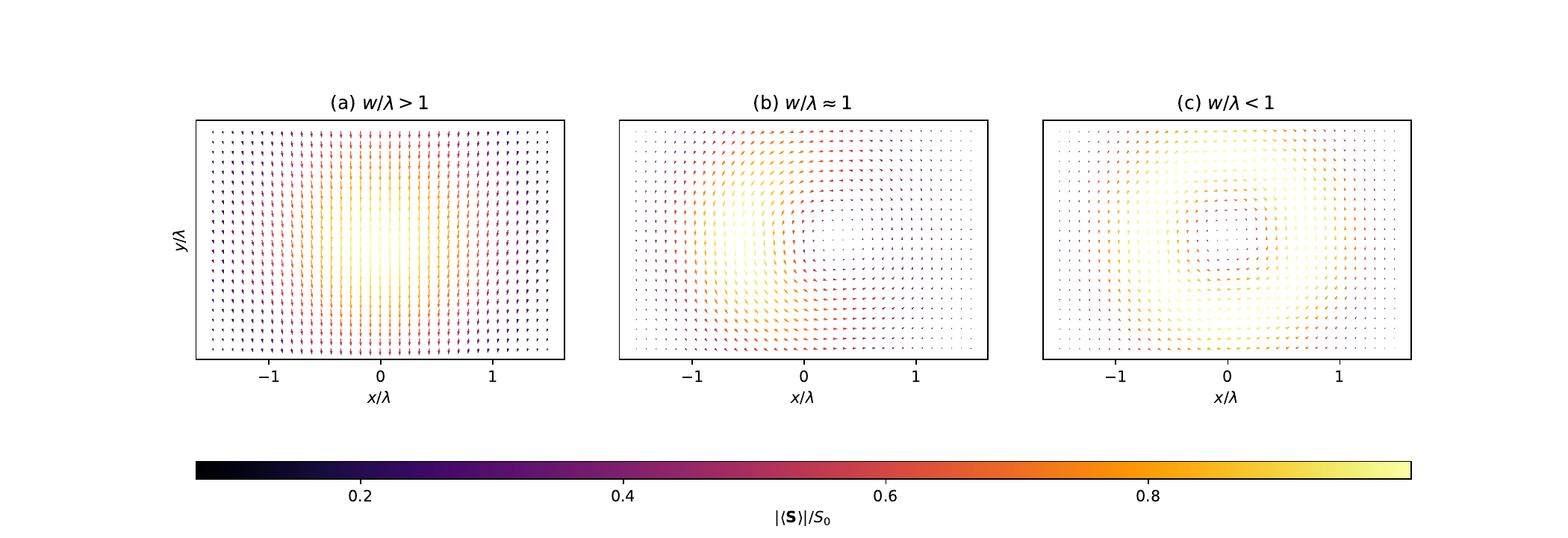}
\caption{Topological transition of the near-field energy flow as a function of normalized groove width.}
\label{fig:topology}
\end{figure}

\begin{figure}[h]\centering
\includegraphics[width=0.7\linewidth]{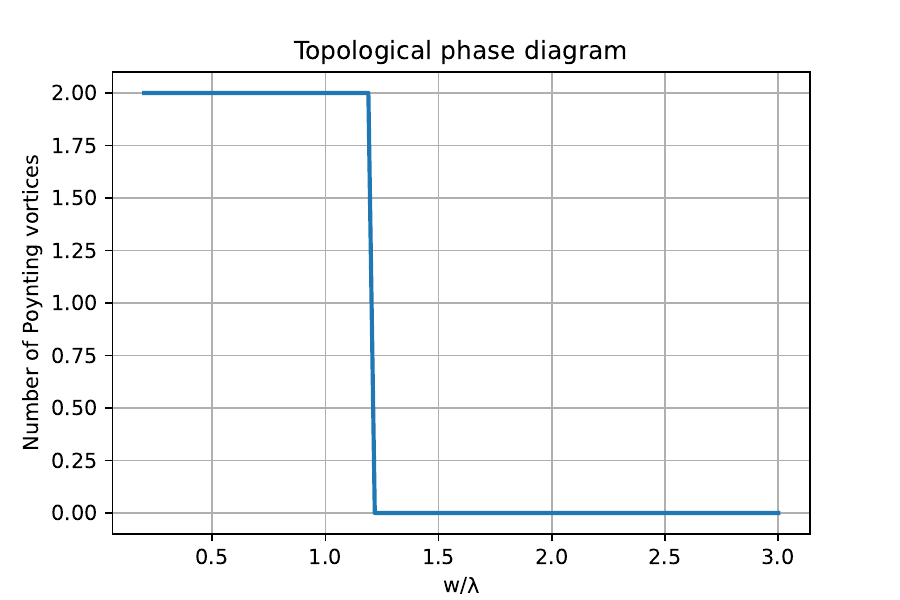}
\caption{ Evolution of the near-field energy-flow topology as a function of the
normalized groove width $w/\lambda$. The diagram summarizes the number and type of
Poynting-vector singularities observed in the near field, revealing discrete
topological regimes separated by the emergence or annihilation of vortex--saddle
pairs.
}
\label{fig:diagram}
\end{figure}

\begin{figure}[h]\centering
\includegraphics[width=\linewidth]{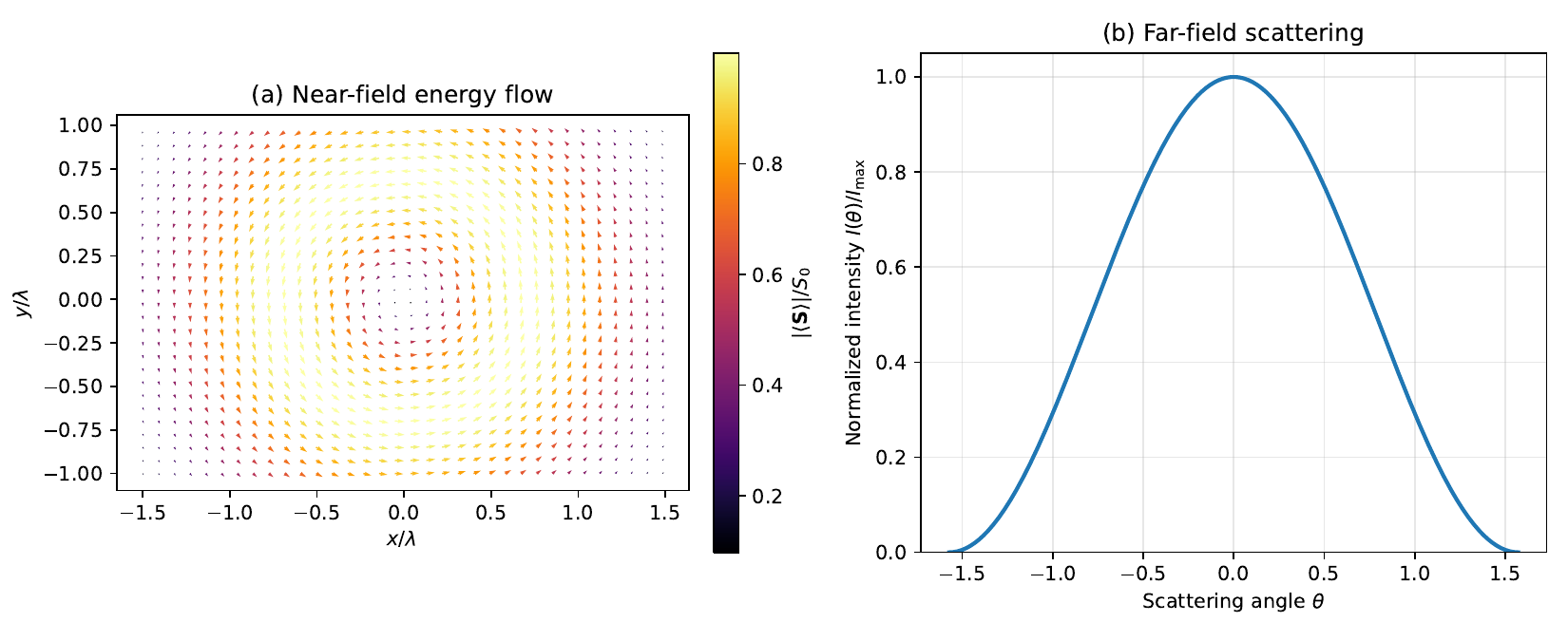}
\caption{ Connection between near-field energy-flow topology and far-field scattering.
(a) Normalized near-field Poynting vector $\langle \mathbf{S} \rangle/S_0$ in the
vicinity of the groove aperture (spatial coordinates normalized to the wavelength),
illustrating the redirection of energy toward the groove axis in the subwavelength
regime. (b) Corresponding far-field scattering pattern shown as normalized angular
intensity $I(\theta)/I_{\max}$, exhibiting a convex profile as a direct consequence
of the near-field energy-flow reorganization.
.}
\label{fig:nearf}

\end{figure}

\bibliographystyle{unsrt}

\end{document}